\documentclass[aps,prl,twocolumn,color,pdflatex]{revtex4-1}
\usepackage{graphicx}% Include figure files
\usepackage{dcolumn}% Align table columns on decimal point
\usepackage{amsmath,amssymb,bbold,bm}
\usepackage{hyperref}
\hypersetup{colorlinks=true,citecolor=black,linkcolor=black,urlcolor=black}
\usepackage{amsfonts}
\usepackage{listings}
\usepackage{braket}
\usepackage{float,latexsym}
\begin{document}

\title{Electrical manipulation of Majorana fermions in an interdigitated superconductor-ferromagnet device}
\author{Shu-Ping Lee$^1$, Jason Alicea$^2$, and Gil Refael$^{1,3}$}
\affiliation{$^1$Department of Physics, California Institute of Technology, Pasadena, CA 91125}
\affiliation{$^2$Department of Physics and Astronomy, University of California, Irvine, CA 92697}
\affiliation{$^3$Dahlem Center for Complex Quantum Systems, Freie Universit\"at Berlin, Arnimallee 14, 14195 Berlin, Germany}
\date{\today}
\begin{abstract}
We show that a topological phase supporting Majorana fermions can form in a 2D electron gas (2DEG) adjacent to an interdigitated superconductor-ferromagnet structure.  An advantage of this setup is that the 2DEG can induce the required Zeeman splitting and superconductivity from a single interface, allowing one to utilize a wide class of 2DEGs including the surface states of bulk InAs.  We demonstrate that the interdigitated device supports a robust topological phase when the finger spacing $\lambda$ is smaller than half of the Fermi wavelength $\lambda_F$.  In this regime, the electrons effectively see a ``smeared" Zeeman splitting and pairing field despite the interdigitation.  The topological phase survives even in the opposite limit $\lambda>\lambda_F/2$, though with a reduced bulk gap.  We describe how to electrically generate a vortex in this setup to trap a Majorana mode, and predict an anomalous Fraunhofer pattern that provides a sharp signature of chiral Majorana edge states.
\end{abstract}
\maketitle

Topological superconductors have attracted considerable recent interest because they may provide the first unambiguous realization of Majorana fermions in any physical setting.  The pursuit of these elusive objects in condensed matter \cite{BeenakkerMajoranaReview,J.A.MajoranaReview} is motivated largely by the non-Abelian statistics \cite{Read-Green-2D-P-wave-superconductor,Ivanov-2001-Non-Abelian,alicea2011non} that they underpin, which is widely sought for quantum computation \cite{RevModPhys.80.1083}. Although much attention recently has focused on finding Majorana fermions in 1D systems \cite{Kitaev-Unpaired-Majorana-1D-wire,Fu-Kane-1D-Fractional-Josephson-2009,PhysRevLett.105.077001,PhysRevLett.105.177002,Topological-Insulator-Nanoribbons, Leo-Kouwenhoven-Majorana-Science}, 2D platforms \cite{Read-Green-2D-P-wave-superconductor,PhysRevLett.100.096407,Sato-Fujimoto-Cold-atom,PhysRevLett.104.040502,Patrick-Lee-P-wave-suerpconductor,PhysRevB.81.125318,SCZhang-Half-Metal} offer some unique virtues such as the ability to perform interferometry \cite{Stern-Halperin-Non-Abelian-Quantum-Hall-State,Bonderson-Kitaev-Shtengel-Non-Abelian-Statistics-Fractional-Quantum-Hall-State, PhysRevLett.102.216403,PhysRevLett.102.216404,JDSau-Detection-Majorana, PhysRevB.83.104513, Eytan-Detect-MajoranaFermion} to probe non-Abelian statistics. One promising 2D scheme involves a quantum well sandwiched between an $s$-wave superconductor and a magnetic insulator \cite{PhysRevLett.104.040502}.  Fabricating this device is, however, rather nontrivial as one must synthesize high quality interfaces on both sides of the quantum well---which is typically buried in a heterostructure.  One can avoid a multilayered architecture by invoking a specific type of 2D electron gas (2DEG) with appreciable Rashba and Dresselhaus coupling \cite{PhysRevB.81.125318}, but the candidate materials for this proposal are limited.

\begin{figure}
\centering
\includegraphics[width=7cm]{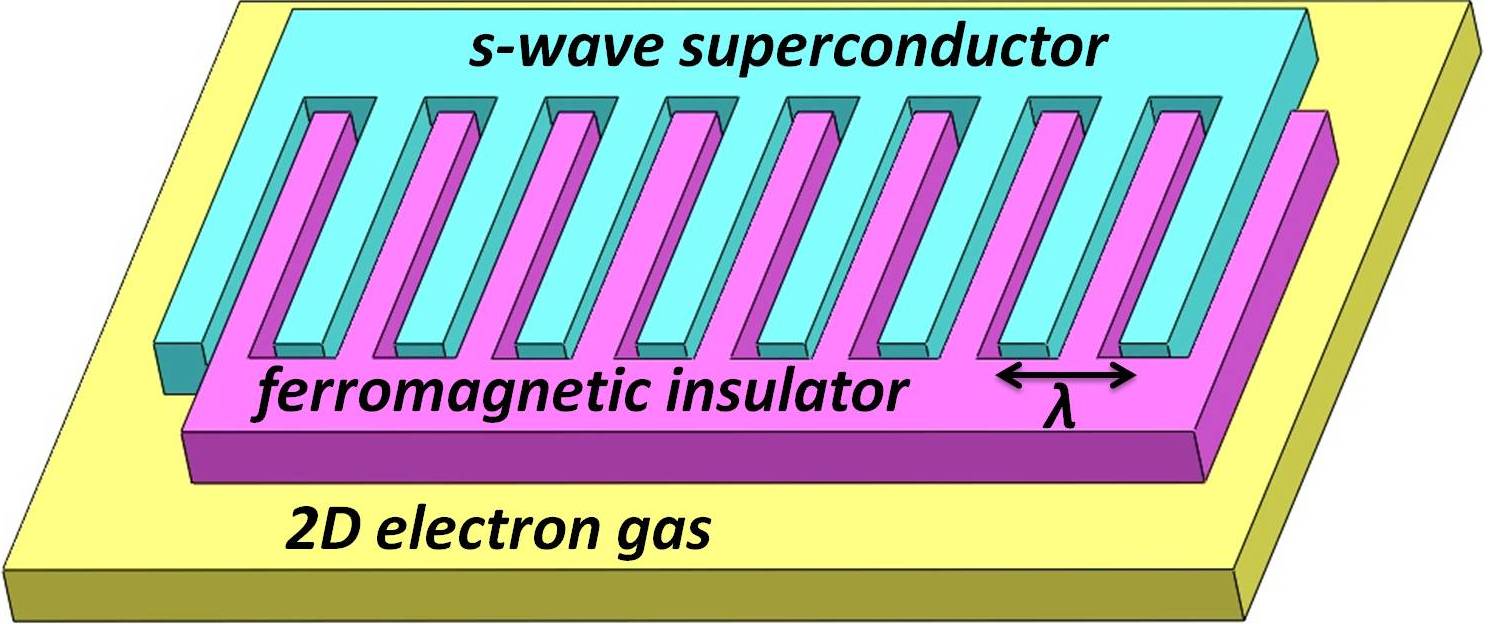}
%\vspace{-10pt}
\caption{Schematic of the proposed interdigitated superconductor-ferromagnet 2DEG architecture. The device supports a robust topological phase when the finger spacing $\lambda$ is smaller than half of the Fermi wavelength (\emph{i.e.} $\lambda<\lambda_F/2$).  A topological phase can also appear in the regime $\lambda>\lambda_F/2$, though generally with a suppressed gap.  Table I provides values of $\lambda_F$ for select 2DEGs when the chemical potential is set to $\mu = 0$.}
\label{SCFMDevice}
\end{figure}

In this manuscript we introduce a new 2D Majorana platform (Fig.\ \ref{SCFMDevice}) consisting of interdigitated superconductor/ferromagnet insulator strips deposited on a 2DEG (periodically modulated 1D topological superconductors were considered in Ref.\ \cite{Avoidance-of-Majorana-Resonances}). The proposed setup exhibits several virtues.  For one, our device requires interface engineering on only one side of the 2DEG---alleviating one experimental challenge with previous semiconductor-based proposals.  Because of this feature one can also employ a wider variety of 2DEGs, including surface states of bulk semiconductors such as InAs \cite{PhysRevB.61.15588,PhysRevB.67.165329,PhysRevB.82.235303}.  Meanwhile, this structure naturally allows one to \emph{electrically} generate vortices to trap Majorana zero-modes, potentially allowing Majorana fermions to be braided using currents similar to the proposal of Ref.\ \cite{Romito-Alicea-Refael-Oppen-Manipulating-Majorana-fermions-using-supercurrents}.  We further show that in our device (as well as any 2D topological superconductor) Majorana edge states can be detected by observing an anomalous shift of the zeros in the Fraunhofer pattern measured in a long Josephson junction.

\begin{table}
\centering
\begin{tabular}{c||c|c|c}
2DEG  & $\alpha$ [eV${\rm \AA}$]  & $m/m_e$  & $\lambda_F$ [$\mu$m]\\
\hline
InGaAs/InAlAs \cite{PhysRevLett.89.046801}& 0.05   & 0.04    & 1.19 \\
InSb/InAlSb \cite{PhysRevB.79.235333}    & 0.14   & 0.0139  & 1.22 \\
Bulk InAs surface \cite{PhysRevB.61.15588,PhysRevB.67.165329,PhysRevB.82.235303}           & 0.11   & 0.03    & 0.72 \\
\end{tabular}
\caption{Effective mass $m$ in units of the electron mass $m_e$, Rashba coupling strength $\alpha$, and Fermi wavelength $\lambda_F$ evaluated at $\mu = 0$ for the 2DEG's listed in the left column.}
\end{table}

We model the semiconductor in this device with the following Hamiltonian,
\begin{eqnarray}
H &=& \int d^2{\bf r}\bigg{\{}\psi^{\dagger} \left[-\frac{\hbar^2\nabla^2}{2m}-\mu-i\alpha(\sigma^x\partial_y-\sigma^y\partial_x) \right]\psi
\nonumber\\
&+& V_z({\bf r})\psi^{\dagger}\sigma^z \psi+
\left[\Delta({\bf r}) \psi_{\uparrow} \psi_{\downarrow}+{\rm H.c.}\right]\bigg{\}},
\label{2}
\end{eqnarray}
where $\psi^\dagger_\sigma$ creates an electron with spin $\sigma$ and effective mass $m$, $\mu$ is the chemical potential, $\alpha$ is the Rashba coupling strength, and $\sigma^{a}$ are Pauli matrices that contract with the spin indices.  The spatially varying Zeeman and pairing fields induced by the alternating ferromagnetic and superconducting strips are respectively denoted by $V_z({\bf r})$ and $\Delta({\bf r})$.  For simplicity we will retain only their maximal Fourier components and take $V_z({\bf r})=2\overline V_z\sin^2(\frac{1}{2}Qx)$ and $\Delta({\bf r})=2\overline \Delta\cos^2(\frac{1}{2}Qx)$.  Here $\overline V_z$ and $\overline\Delta$ are the spatial average of these quantities, which modulate at wavevector $Q=2\pi/\lambda$ with $\lambda$ the finger spacing shown in Fig.\ \ref{SCFMDevice}.  This choice is expected to not only quantitatively capture the effects of interdigitation, but as we will see also leads to an intuitive physical picture for the device's behavior.

As a primer it is worth recalling the physics of the sandwich structure originally proposed by Sau \emph{et al.} \cite{PhysRevLett.104.040502}, where a uniform Zeeman field $V_z^{\rm unif}$ opens a chemical potential window in which only one Fermi surface is present.  Incorporating $s$-wave pairing with strength $\Delta^{\rm unif}$ in this regime effectively drives the 2DEG into a topological $p+ip$ superconductor due to the interplay with spin-orbit coupling. \cite{PhysRevLett.104.040502,PhysRevB.81.125318,J.A.MajoranaReview} Quantitatively, the topological phase appears provided $(V_z^{\rm unif})^2>(\Delta^{\rm unif})^2+\mu^2$.  In our interdigitated setup it is natural to expect that when the Fermi wavelength $\lambda_F$ for the semiconductor greatly exceeds the finger spacing $\lambda$, electrons in the 2DEG effectively experience `smeared' Zeeman and pairing fields with strength $\overline V_z$ and $\overline \Delta$.  Similar physics to the uniform case ought to then emerge---in particular, a topological phase when $\overline V_z^2 \gtrsim \overline \Delta^2+\mu^2$.

\begin{figure}
\includegraphics[width=8cm]{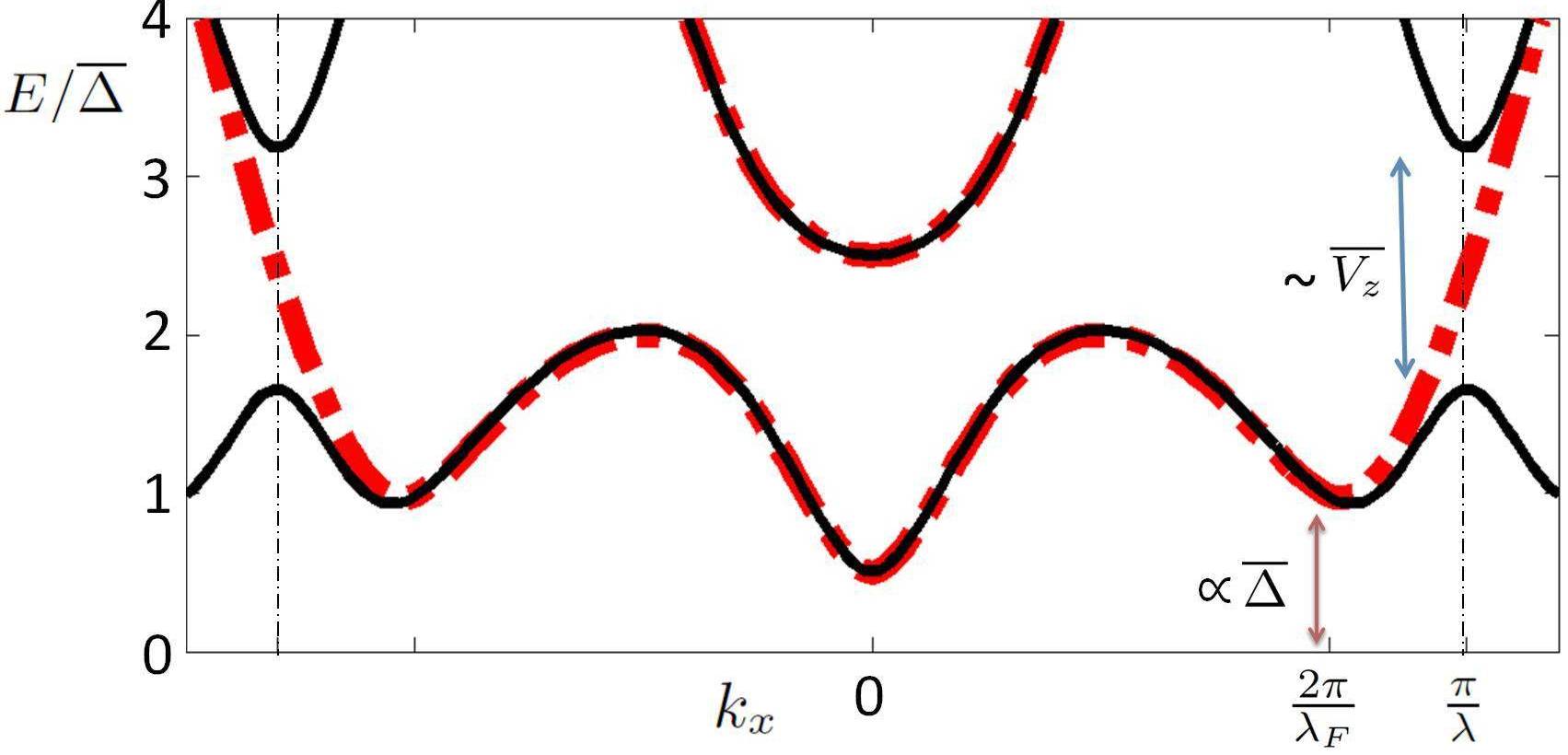}
\caption{Bulk quasiparticle spectrum versus $k_x$ in a uniform structure (dashed curve) and interdigitated device with $\lambda_F/\lambda=2.6$ (solid curve).  In both cases we use parameters $k_y = 0$, $\mu=0$, $\overline{V_z}=1.5\overline{\Delta}$, and $m\alpha^2=3\overline{\Delta}$.
The red arrow indicates the pairing gap at the Fermi momentum $k_F\equiv\frac{2\pi}{\lambda_F}\approx2m\alpha/\hbar^2$, while the blue arrow denotes the degeneracy gap opened at $k_x=\frac{\pi}{\lambda}$ due to the interdigitation.  Note that the excitation spectra for the uniform and interdigitated systems differ appreciably only at rather higher energies here.}
\label{BulkBdG}
\end{figure}

To confirm this intuition and extract the phase diagram for arbitrary $\lambda_F/\lambda$, we study the quasiparticle spectrum for Eq.\ (\ref{2}). Defining a Nambu spinor $\Psi_{\bf k}=[\psi_{\uparrow}({\bf k}),\psi_{\downarrow}({\bf k}),\psi^{\dagger}_{\downarrow}(-{\bf k}),-\psi^{\dagger}_{\uparrow}(-{\bf k})]^{T}$, the Hamiltonian can be written in momentum space as
\begin{eqnarray}
  H&=&\int \frac{d^2{\bf k}}{(2\pi)^2} (\overline{\mathcal{H}}_{\bf k}+\delta \mathcal{H}_{\bf k})
  \label{MomentumBdGeq2}
  \\
  \overline{\mathcal{H}}_{\bf k}&=&\Psi_{\bf k}^{\dagger} \bigg{[}\left(\frac{\hbar^2k^2}{2m}-\mu\right)\tau^z+\alpha\left(k_y\sigma^x-k_x\sigma^y\right)\tau^z
  \nonumber \\
  &+& \overline V_z\sigma^z+\overline \Delta\tau^x\bigg{]}\Psi_{\bf k}
   \\
  \delta \mathcal{H}_{\bf k}&=&\Psi_{\bf k}^{\dagger} \left[-\frac{\overline V_z}{2} \sigma^z+\frac{\overline \Delta}{2}\tau^x\right]\Psi_{{\bf k} + Q{\bf \hat{x}}}+{\rm H.c.}
\end{eqnarray}
with $\tau^a$ Pauli matrices that act in particle-hole space.  The Hamiltonian $\overline{\mathcal{H}}_{\bf k}$ describes a semiconductor proximate to a uniform superconductor and ferromagnet and is precisely the model studied in Ref.\ \cite{PhysRevLett.104.040502}.  The bulk excitation spectrum obtained from $\overline{\mathcal{H}}_{\bf k}$ in the topological phase with $\mu=0$, $\overline{V_z}=1.5\overline{\Delta}$, $m\alpha^2=3\overline{\Delta}$ and $k_y = 0$ appears in the red dashed lines of Fig.\ \ref{BulkBdG}; roughly, the gap at $k_x = 0$ is set by $\overline{V}_z$ while $\overline\Delta$ determines the gap at the Fermi wavevector $k_F = 2\pi/\lambda_F$.  Our interdigitated structure produces a new term $\delta \mathcal{H}_{\bf k}$ that couples spinors with wavevectors ${\bf k}$ and ${\bf k} \pm Q{\bf \hat{x}}$.  As we `turn on' these couplings the spectrum of $\overline{\mathcal{H}}_{\bf k}$ evolves very similarly to the band structure of free electrons in a weak periodic potential \cite{ashcroft1976solid}.  In particular, the dominant effect of $\delta \mathcal{H}_{\bf k}$ is to open a gap in the excitation spectrum whenever the energies cross Bragg planes at $k_x = \pm Q/2 = \pm\pi/\lambda$ (modulo reciprocal lattice vectors).  For momenta away from these values $\delta \mathcal{H}_{\bf k}$ couples states that are far from resonant and hence perturbs these only weakly.

\begin{figure}
\includegraphics[width=8.6cm]{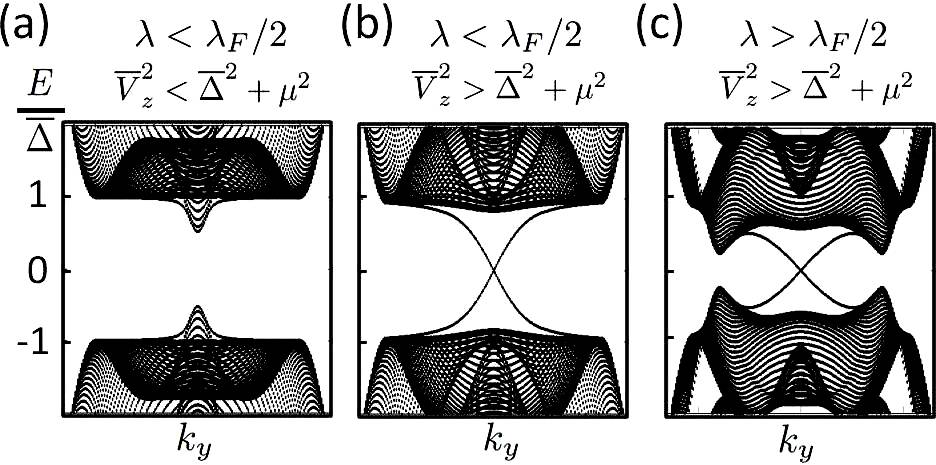}
\caption{Quasiparticle spectrum in various regimes for an interdigitated device with periodic boundary conditions along $y$ but open boundary conditions along $x$.  In all parts we take $\mu=0$ and $m\alpha^2=3\overline{\Delta}$, while the finger spacing and Zeeman energy vary as (a) $\lambda_F/\lambda = 2.5$, $\overline{V_z}=0.5\overline{\Delta}$, (b) $\lambda_F/\lambda = 2.5$, $\overline{V_z}=2\overline{\Delta}$, and (c) $\lambda_F/\lambda = 1.5$, $\overline{V_z}=2\overline{\Delta}$.  A trivial state appears in (a) while the larger Zeeman field in (b) drives a topological phase supporting chiral Majorana edge states within the bulk gap.  Interestingly, the topological phase and associated Majorana edge states survive even in (c) despite the relatively small ratio of $\lambda_F/\lambda$.}
\label{BdG}
\end{figure}

It follows that for $\lambda_F/\lambda \gg 1$ the periodic modulation modifies the quasiparticle spectrum appreciably only at very high energies.  This point is illustrated by the solid curves in Fig.\ \ref{BulkBdG}, which display the numerically obtained spectrum for the full Hamiltonian in Eq.\ (\ref{MomentumBdGeq2}) in a repeated zone scheme,
using the same parameters as above but now with $\lambda_F/\lambda=2.6$.  Even for this ratio of $\lambda_F/\lambda$, the spectrum is nearly identical to that of the uniform case away from $k_x = \pm \pi/\lambda$.  When $\lambda_F/\lambda \gg 1$ one can clearly incorporate $\delta\mathcal{H}_{\bf k}$ while essentially leaving the bulk excitation gap exhibited by the uniform system intact.  Thus by adiabatic continuity our interdigitated device supports a topological phase in this limit provided $\overline V_z^2 \gtrsim \overline \Delta^2+\mu^2$, consistent with the intuition provided earlier.  As further evidence, Figs.\ \ref{BdG}(a) and (b) display the quasiparticle spectrum as a function of $k_y$ in a system with open boundary conditions along the $x$ direction. The data correspond to $\mu=0$, $m\alpha^2=3\overline{\Delta}$ and $\lambda_F/\lambda=2.5$, while the Zeeman energy changes from $\overline{V_z}=0.5\overline{\Delta}$ in (a) to $\overline{V_z}=2\overline{\Delta}$ in (b).  In (a) a trivial gapped state clearly emerges due to the weak Zeeman energy.  The larger $\overline{V}_z$ value in (b), however, satisfies our topological criterion, and one indeed sees the signature gapless chiral Majorana edge states inside of the bulk gap.

\begin{figure}
\centering
\includegraphics[width=8cm]{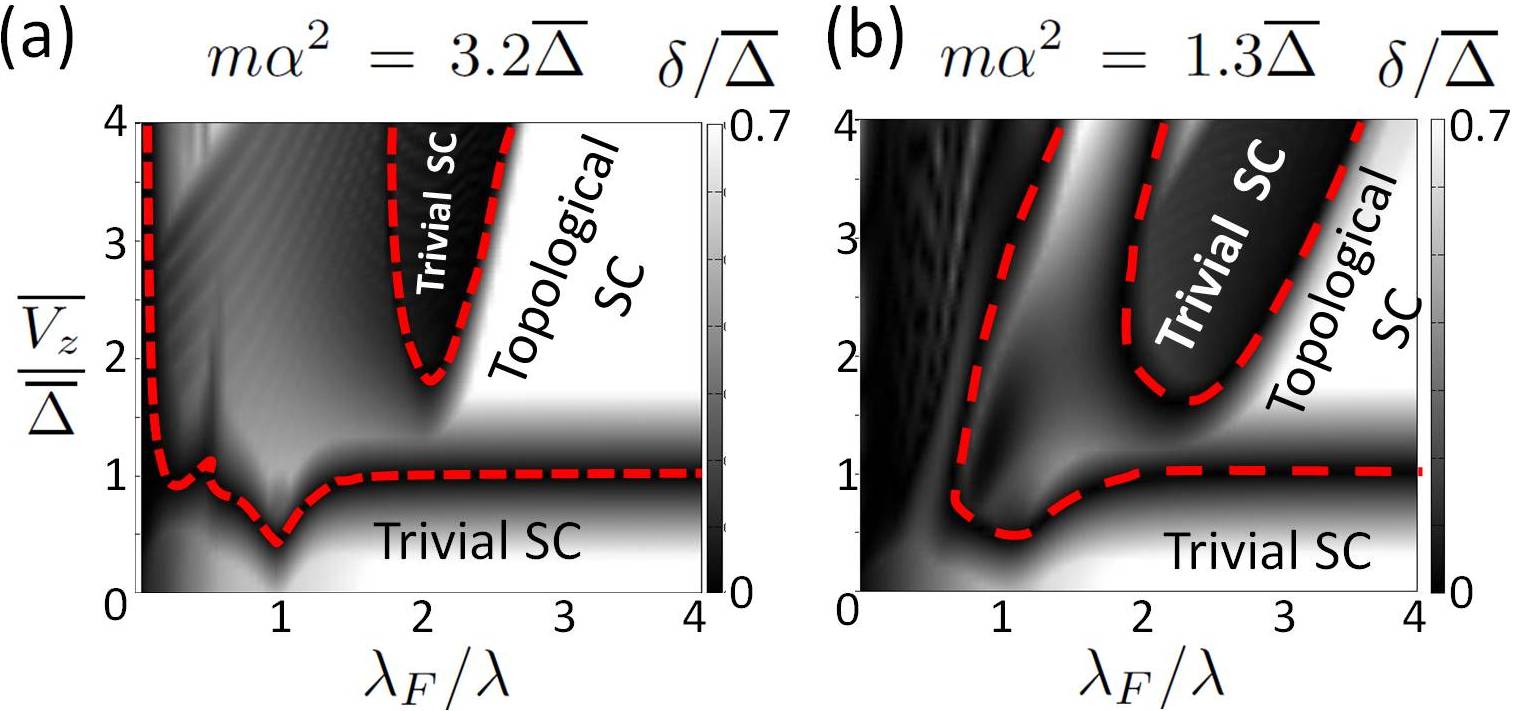}
\caption{Phase diagrams for $\mu = 0$ and spin-orbit energies (a) $m\alpha^2=3.2\overline{\Delta}$ and (b) $m\alpha^2=1.3\overline{\Delta}$.  The horizontal axis represents ratio of the Fermi wavelength $\lambda_F$ to the finger spacing $\lambda$, while the vertical axis is the Zeeman energy normalized by pairing strength.  The shading indicates the bulk gap $\delta$ normalized by $\overline{\Delta}$. Red dashed lines denote the boundary between topological phase and trivial phases.  }
\label{PhaseDiagram2}
\end{figure}

As one reduces the ratio $\lambda_F/\lambda$ to a value of order one or smaller, the physics becomes considerably more subtle.  Indeed, once $\lambda_F/\lambda \approx 2$ the Bragg plane at $k_x = Q/2$ approaches the Fermi wavevector, and the pairing gap can then be dramatically altered by the interdigitation.  We ascertain the global phase diagram of our device by numerically computing the minimum excitation gap $\delta$ for a system on a torus as a function of $\lambda_F/\lambda$ and $\overline{V}_z/\overline\Delta$.  Figure \ref{PhaseDiagram2} shows the results for $\mu=0$ and spin-orbit energies of $m\alpha^2=3.2\overline{\Delta}$ in (a) and $m\alpha^2=1.3\overline{\Delta}$ in (b).  The following points are noteworthy here: 1) At `large' $\lambda_F/\lambda$ topological superconductivity appears when $\overline{V_z} \gtrsim \overline\Delta$, in line with our results above.  2) The topological phase survives over a range of parameters even for rather small values of $\lambda_F/\lambda$, though the gap is generally reduced compared to the large $\lambda_F/\lambda$ limit.  Figure \ref{BdG}(c) illustrates the spectrum in the $\lambda_F/\lambda<2$ regime for a system with open boundary conditions along $x$; just as in Fig.\ \ref{BdG}(b) the characteristic chiral edge states again appear here. 3) Interestingly, for $\lambda_F/\lambda\sim 1$ the critical value of $\overline{V}_z$ required to generate the topological phase \emph{decreases} compared to the uniform case.

\begin{figure}
\centering
\includegraphics[width=8.5cm]{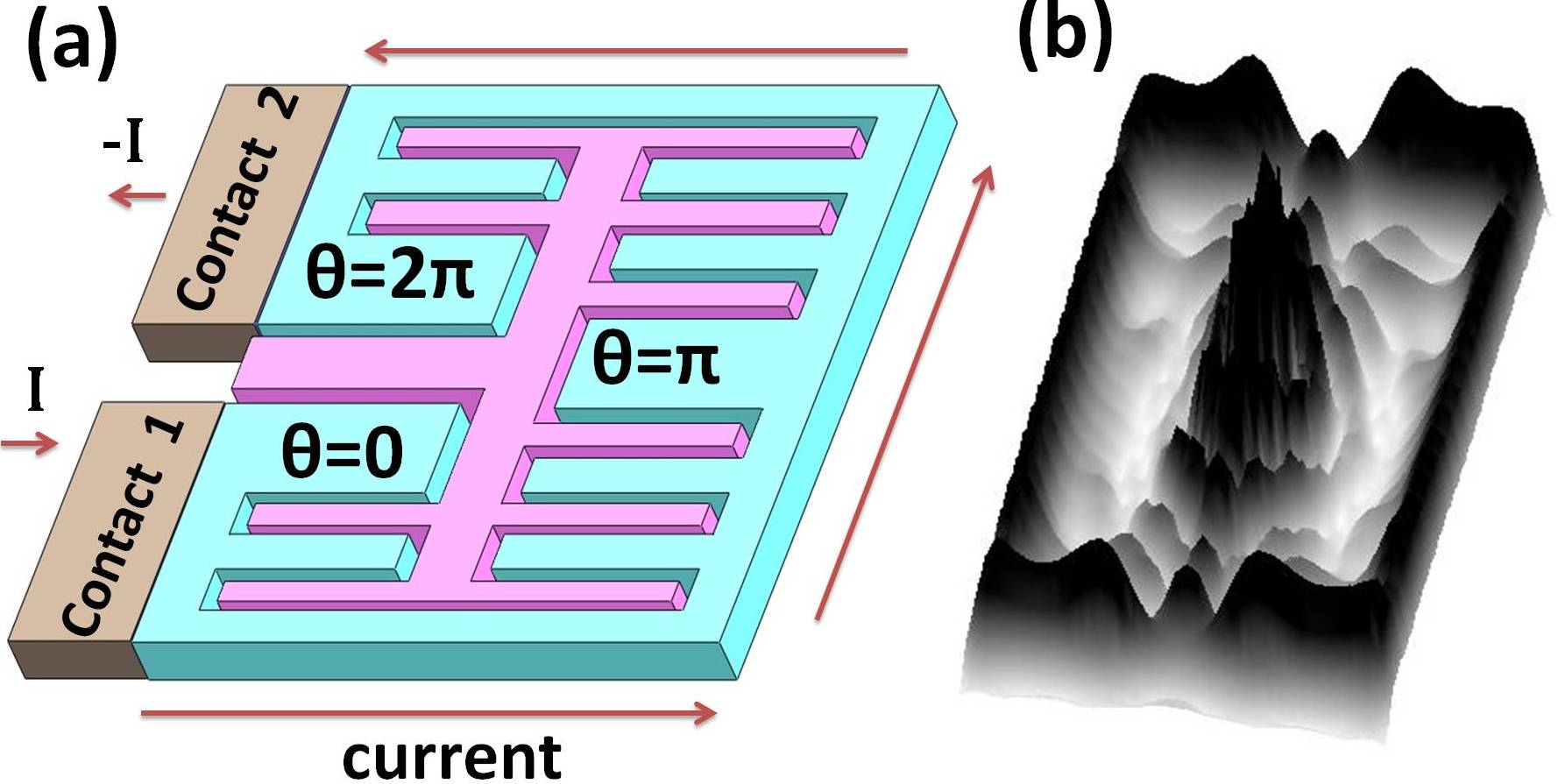}
\caption{(a) Scheme to electrically stabilize a vortex binding a Majorana zero-mode.  Here the singular phase winding is induced by current flowing from contact 1 to contact 2, rather than from a magnetic field.  In (b) we illustrate the probability density extracted from the near-zero energy mode generated by a current-induced vortex at the center of the device (parameters are $\mu=0$, $m\alpha^2=1.3\overline{\Delta}$, $\overline{V}_z=2\overline{\Delta}$, and $\lambda=\lambda_F/4$).  The large central peak corresponds to the Majorana bound to the vortex, which hybridizes weakly with the outer Majorana running along the perimeter.  }
\label{TunnelingCurrent}
\end{figure}

Having numerically demonstrated that our device exhibits a topological phase with an edge state, we now describe how the interdigitated structure naturally allows us to \emph{electrically} generate vortices to trap Majorana zero-modes. Consider the setup of Fig.\ \ref{TunnelingCurrent}.  Supercurrent flowing from contact 1 to contact 2 produces a winding in the superconducting phase $\theta({\bf r})$ across the fingers in the device.
When the phase difference between the contacts approaches $2\pi$, a vortex forms near the center of the system to minimize the energy $E\propto\int d^2{\bf r}(\nabla\theta)^2$.  The spatial profile of the phase follows from the supercurrent $ {\bf j}({\bf r})\propto\Delta^*({\bf r})\nabla\Delta({\bf r})-\Delta({\bf r})\nabla\Delta^*({\bf r})$, with $\Delta({\bf r})=\Delta_{SC}({\bf r})e^{i\theta({\bf r})}$, where the pairing potential's magnitude satisfies $\Delta_{SC}({\bf r})=2\overline{\Delta}$  beneath the superconductors [blue regions in Fig.\ \ref{TunnelingCurrent}(a)] and goes to zero under the ferromagnets [pink regions in Fig.\ \ref{TunnelingCurrent}(a)]. In particular, one can extract $\theta({\bf r})$ by iterating the current conservation equation $\nabla\cdot{\bf j}({\bf r})=0$ subject to boundary conditions along the system's perimeter. With this phase in hand, one can diagonalize the Hamiltonian in the presence of a current-induced vortex and extract  the wavefunctions for each quasiparticle state.  Figure \ref{TunnelingCurrent}(b) illustrates the resulting probability distribution for the near-zero-energy state in the spectrum; the large central peak corresponds to a localized Majorana mode bound to the vortex core, while the outer peak represents a second Majorana mode running along the edge.

\begin{figure}
\includegraphics[width=8.5cm]{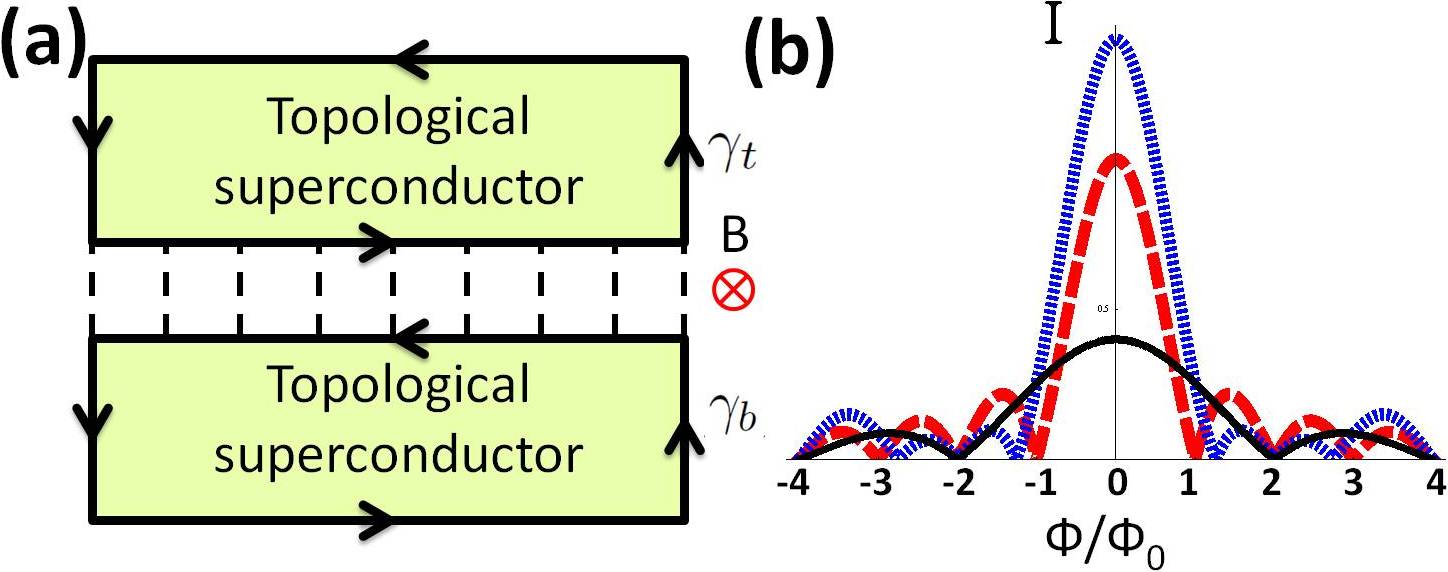}
\caption{(a) Long Josephson junction formed by adjacent topological superconductors. A magnetic field $\vec{B}$ orients perpendicular to the plane and uniformly penetrates through the junction.  (b) The solid black curve represents the magnitude of the tunneling current arising from coupled Majorana zero-modes at the edge.  This contribution exhibits zeros at even multiples of the flux quantum in sharp contrast to the Fraunhofer pattern exhibited by ordinary $s$-wave superconductor junctions (red dashed curve). The blue curve represents the anomalous Fraunhofer pattern that would arise in an experiment due to the Majorana-mediated component and a parallel conventional current contribution.}
\label{FraunhoferPatternForMajorana}
\end{figure}

Finally, we discuss the detection of Majorana edge states in the topological phase exhibited by our device (or, equivalently, any other realization) via an unconventional Fraunhofer pattern. \footnote{This idea has been independently proposed by Julia Meyer and Manuel Houzet.}  Consider a pair of topological superconductors forming a long Josephson junction of width $w$ pierced by a magnetic field [see Fig.\ \ref{FraunhoferPatternForMajorana}(a)].  At low energies it suffices to focus only on the chiral edge states, which can be modeled by an effective Hamiltonian  $H = H_t + H_b + H_{\rm tunneling}$. \cite{Read-Green-2D-P-wave-superconductor,PhysRevB.75.045317, Eytan-Detect-MajoranaFermion}  The first two terms $H_{t/b}=\pm iv\hbar\int dx \gamma_{t/b}\partial_x \gamma_{t/b}$ describe the kinetic energy for the top/bottom edge states, with $\gamma_{t/b}$ Majorana operators and $v$ the edge velocity.  The last term incorporates inter-edge tunneling with strength $t$ at the interface and reads $H_{\rm tunneling}=it\int_{-w/2}^{w/2} dx \gamma_t\gamma_b\cos[\theta(x)/2]$, where $\theta(x)$ is the local superconducting phase difference across the junction induced by the applied field.  Neglecting the magnetic field that is produced from the tunneling current, $\theta(x)$ is determined by the external magnetic flux $\Phi$ according to $\theta(x)=\theta_0+2\pi\frac{\Phi}{\Phi_0}\frac{x}{w}$ ($\Phi_0$ is the flux quantum and $\theta_0$ is the phase difference at the junction's center).

The Majorana-mediated contribution to the local current density flowing across the junction follows from $j(x)=\frac{et}{\hbar}\sin[\theta(x)/2]i\gamma_t\gamma_b$.  We calculate the current perturbatively in $t$ assuming the weak-tunneling limit $\frac{tw}{2\pi\hbar v}<1$ where the hybridization energy is smaller than the level spacing.  In this case the physics depends sharply on whether, at $t = 0$, Majorana zero-modes exist at each edge.  If neither edge supports a zero-mode, then the current vanishes to first order in $t$.  However, if zero-modes exist at both edges (due to an odd number of vortices in their bulk) then a finite current
$\langle j(x)\rangle=\frac{et}{\hbar L}\sin[\theta(x)/2]$ emerges, where $L$ is the superconductors' perimeter. Integrating over the junction width yields a total Majorana-mediated current
\begin{equation}
  I_M=\frac{twe}{\hbar L}\sin\left(\frac{\theta_0}{2}\right)\left[\frac{\sin(\frac{\pi}{2}\frac{\Phi}{\Phi_0})}{\frac{\pi}{2}\frac{\Phi}{\Phi_0}}\right].
\end{equation}
The solid black curve in Fig.\ \ref{FraunhoferPatternForMajorana}(b) illustrates $|I_M|$ as a function of $\Phi$; remarkably, the zeros occur at \emph{even} multiples of $\Phi/\Phi_0$ in contrast to the conventional Fraunhofer pattern shown for comparison in the red dashed curve.  For a sample of size $5\mu m\times 5 \mu m$ with the coupling energy $t=0.025$meV \cite{Eytan-Detect-MajoranaFermion}, we estimate that the typical magnitude of $I_M$ is $\sim 1.5$nA. This result is valid when the edge velocity obeys $v>3\times10^4 m/s$ so that the weak tunneling limit is satisfied. It is important to keep in mind, however, that the experimentally observed current will not be given by $I_M$ alone---a potentially much larger conventional current $I_s$ flows in parallel.  The magnitude of the total current $I_{\rm tot} = I_s + I_M$ is sketched by the blue curve in Fig.\ \ref{FraunhoferPatternForMajorana}(b).  One can infer the existence of $I_M$ by the unconventional Fraunhofer pattern that exhibits shifted zeros as shown in the figure.  We note that very recently an experiment of this type has been performed in a long Josephson junction formed at the surface of a 3D topological insulator \cite{DGG-FraunhoferPattern}, though the findings are rather different from what we predict here.

In conclusion, we have shown that our interdigitated structure exhibits a topological phase that is particularly robust when the finger spacing is smaller than half of the Fermi wavelength.  There the bulk gap can be comparable to that in a uniform system; furthermore, additional perturbations induced by the interdigitation (such as variations in chemical potential and Rashba strength) should play a minor role.  To access this regime the finger spacing should be $\lesssim 600$nm for the quantum wells listed in Table I and $\lesssim 400$nm for the surface state of bulk InAs. We also note that since electrons effectively see `smeared' fields in this limit, the specific interdigitated pattern studied here is by no means required---similar physics should arise, \emph{e.g.}, in checkerboard arrangements.  An interesting feature of our setup is that vortices can be generated by applying currents.  This mechanism may eventually provide a practical means of manipulating and braiding vortices for quantum computation. We also pointed out that chiral Majorana edge states produce an anomalous Fraunhofer pattern that can be observed in any realization of topological $p + ip$ superconductivity.

\begin{acknowledgments}
We are indebted to Charles M. Marcus for proposing the interdigitated structure studied here, and to Julia Meyer for discussions on the Fraunhofer pattern.  We also thank E. Grosfeld, K. T. Law, J. Eisenstein, P. T. Bhattacharjee, S. Iyer, and D. Nandi for illuminating discussions. We are grateful to the Packard Foundation (GR), Humboldt Foundation and to the Institute for Quantum Information and Matter, an NSF Physics Frontiers Center with support of the Gordon and Betty Moore Foundation. JA gratefully acknowledges funding from the National Science Foundation through grant DMR-1055522 and the Alfred P. Sloan Foundation.
\end{acknowledgments}
%\bibliography{InterdigitatedReference3}


\end{document}